\documentclass[10pt]{revtex4}
\usepackage{amssymb}
\usepackage{latexsym}
\usepackage{epsfig}

\usepackage{amsmath}
\begin{document}
\title{New Tsallis agegraphic Dark Energy in Fractal cosmology }
\author{M. Abdollahi Zadeh\footnote{
mazkph@gmail.com}}
\address{Department of Physics, Faculty of Science, Salman Farsi University of Kazerun, Kazerun, Iran}

\begin{abstract}
In this paper, we study the new Tsallis agegraghic dark energy model in the context of fractal cosmology. The different cosmic implications such as statefinder diagnosis, cosmological plane and squared of sound speed in both interacting and non-interacting
scenarios are investigated. The study shows
that the equation of state (EoS) parameter gives quintessence-like nature of Universe in both interacting and non-interacting cases. The results of the stability analysis for the distinct estimations of the nonadditive parameter $\delta$ and coupling constant $b^2$ indicate that this model is unstable. Moreover, $\omega_D-{\omega}^{\prime}_{D}$ plane is in the freezing region in other words, cosmic expansion is more accelerating in the framework of fractal spacetime.
\end{abstract}
\maketitle

\section{Introduction}

  About 6 Gyr ago, a strange thing happened. The Universe that till before this date had the decelerated expansion, with a change of rhythm, entered the accelerated phase. Because the Einstein field equations do not predict this event, the efforts to understand this issue intensified. The first step was to add a cosmological constant to the Einstein\textquoteright s equation of General Relativity. However, the $\Lambda$ cosmology is consistent with the cosmological observations, the problem of amount and origin of it \cite{p5}, led to the search of alternative candidates as the next steps. The steps that can be summarized as follows, to introduce various dynamical dark energy models, keeping the Einstein field equations as the gravitational theory or to modify the standard theory of gravity on extragalactic scales \cite{Riess,Riess1,Riess2}. Of course, we also have a missing in the Universe that is responsible for the formation of cosmic structures. This missing, called dark matter,  together with dark energy, comprise approximately $95\%$ of our Universe.

One of the important fields of research in theoretical physics is to formulate a consistent theory of quantum gravity. The efforts such as loop quantum gravity to Horava-Lifshitz theory \cite{Hora,Hora1} are the results of this hand. All these theories give a hint that the spectral dimension of spacetime at the Planck scale is 2, while it is four-dimensional at the large scales \cite{Hora2}. It shows that ordinary geometry is inadequate to describe the microscopic texture of spacetime and we should find the model that its dimension changes with the scale of the Universe. This is a characteristic of fractals and refers to the fact that, the  application of quantum mechanics to spacetime leads to a fractal geometry. After that Horava proposed his theory \cite{Hora,Hora1}, which suffered from some problems, Calcagni \cite{Calcagni2,Calcagni} introduced a model of quantum gravity in a fractal Universe that was Lorentz-invariant and free from ultraviolet divergence and discussed its implications for cosmology. Because at late times, an imprint of the nontrivial short-scale geometry might survive as a cosmological constant \cite{Calcagni}, the fractal effects in the fractal cosmology may be another approach to cosmic acceleration or the sources of dark energy.

The New Agegraphic Dark Energy (NADE) model is one of the dark energy models that keeps the structure of Einstein\textquoteright s equations. This model originates from the uncertainty relation of quantum mechanics, where the conformal time $\eta$, is IR cutoff. In fact, this model discards the difficulty to describe the matter-dominated epoch in the Agegraphic Dark Energy model (ADE). Since this approach is based on the entropy relation, each modification to the system entropy may change the ADE model \cite{Cai, Wei, Wei1}.
 
Entropy is an additive function when we study statistically independent systems. But in gravitational or cosmological ones, where there is long range interaction, the energy between the different parts of the system is not negligible compared with the total energy, thus the entropy should be generalized to a nonadditive one. Based on this argument, Tsallis and Cirto introduced a new entropy for the black hole as $ S_{T}=\gamma A^{\delta}$, \cite{THDE2}  which can be generalized to cosmological cases. Here, $\delta$ denotes the nonadditive parameter.

Recently, inspired by the Tsallis generalized entropy, Tsallis ADE (TADE) models by using the age of the Universe and the conformal time as the IR cutoffs, have been introduced \cite{TNADE}. Ghaffari \textit {et al} \cite{fractal} have studied cosmological consequences of Tsallis holographic dark energy model (THDE) considering the Hubble radius as the IR cut-off in the framework of fractal spacetime. The study of the dynamics of a fractal Universe filled with the THDE by assuming an interacting case has been studied by Al Mamon \cite{fractal1}. Debnath \textit {et al} \cite{fractal2} have investigated the non-canonical scalar field model in the background of fractal spacetime. Jawad \textit {et al} \cite{fractal3} discussed the cosmological implications of pilgrim dark energy model in fractal cosmology. More works, in the background of fractal cosmology also can be found in Refs \cite{Karami0,Karami1,Karami2,Karami3,Karami4,Karami5,Karami6,Karami7}. In the present work, we study the NTADE model in the framework of the fractal Universe. The paper organized as follows: we study the fractal cosmology and introduce some of the cosmological parameters for distinguishing among DE models in section II. In section III, we present the cosmological behavior of the NTADE in the framework fractal spacetime without an interaction term. Also, the features of this model considering interaction term are studied in section IV. Finally, we summarize the results of the present work in the last section.
\section{Fractal cosmology}

Assuming a fractal spacetime, the total action of Einstein gravity
($S\equiv S_G+S_m$), is decomposed as \cite{Calcagni,Calcagni2}
 \begin{equation}
   S_G=\frac{1}{16\pi G}\int
   d\varrho\left(x\right)\sqrt{-g}\left(R-2\Lambda-\omega\partial_{\mu}\nu\partial^{\mu}\nu\right),\
   \end{equation}
   and
   \begin{equation}
  S_m=\frac{1}{16\pi G}\int d\varrho\left(x\right)\sqrt{-g}\mathcal{L}_m,\
  \end{equation}
  where $S_G$ and $S_m$ are the gravitational and matter parts of
  action, respectively. 
Also, $\nu$ and $\omega$ stand for
the fractional function and fractal parameter, respectively.
Moreover, $d\varrho\left(x\right)$ denotes the Lebesgue-Stieltjes
measure generalizing the standard four-dimensional measure $d^4 x$.

In a fractal universe, the Friedmann equation can be obtained by
taking the action variation with respect to $g_{\mu\nu}$ as
  \begin{equation}\label{frwfractal}
  H^2+\frac{k}{a^2}+H\frac{\dot{\nu}}{\nu}-\frac{\omega}{6}\dot{\nu}^2=\frac{1}{3}(\rho_m+\rho_D)+\frac{\Lambda}{3},
   \end{equation}
here, $k$ is the curvature
parameter and $\rho_m$ and $\rho_D$ are also the energy densities of DM
and DE, respectively. Since we are eager to model the accelerated
universe without considering the unknown and mysterious source
$\Lambda$, we will consider $\Lambda=0$, and also, motivated by the
WMAP data, we limit ourselves to $k=0$ (flat universe).

The continuity equation is decomposed as
\begin{eqnarray}\label{conm}
&&\dot{\rho}_m+\left(3H+\frac{\dot{\nu}}{\nu}\right)\rho_m=Q,\\
&&\dot{\rho}_D+\left(3H+\frac{\dot{\nu}}{\nu}\right)\left(\rho_D+P_D\right)=-Q,\label{conD}
\end{eqnarray}
where $P_D$ is the pressure of DE, and $Q$ describes the
interaction term between DE and DM. Here, we consider the $Q=3
b^2 H \rho_m$ case, that $b^2$ is a coupling constant. Now,
assuming the fractional function $\nu=a^{-\gamma}$, the Friedmann
equation and continuity equations can be written as
  \begin{equation}\label{fractal}
   H^2\left(1-\gamma-\frac{\gamma^2\omega
   a^{-2\gamma}}{6}\right)=\frac{1}{3}\left(\rho_D+\rho_m\right),
   \end{equation}
  \begin{equation}\label{fractal1}
  \dot{\rho}_m=\left(3\left(b^2-1\right)+\gamma\right)H\rho_{m_0}a^{3\left(b^2-1\right)+\gamma},
   \end{equation}
    \begin{equation}\label{fractal2}
  \dot{\rho}_D=-\left(3-\gamma\right)\left(1+\omega_D\right)\rho_DH-3b^2H\rho_{m_0}a^{3\left(b^2-1\right)+\gamma},
   \end{equation}
    here, $\omega_D\equiv\frac{P_D}{\rho_D}$, and $\rho_{m_0}$ denotes the present value of DM density.

In all dark energy models, we are dealing with the Hubble parameter $H$ as the expansion rate of the Universe and the deceleration paremeter $q$ as the rate of the acceleration or deceleration of the Universe expansion, where the first case is always positive and the second case is negative. Thus, Sahni \textit {et al}  \cite{Sahni}  introduced a geometrical diagnostic that is model independent and is constructed from a spacetime metric directly, for distinguishing among DE models. 
The geometrical statefinder parameters are defined as 
\begin{equation}\label{rr2}
r=2q^2+q-\frac{\dot q}{H},
\end{equation}
\begin{equation}\label{statefinder}
s=\frac{r-1}{3(q-1/2)}.
\end{equation}

The diffrrent DE models can be identified by $r-s$ plane, for example a fixed point $\{r,s\}=\{1,0\}$ shows $\Lambda$CDM model and $\{r,s\}=\{1,1\}$ describes CDM limit.
Also, another useful method is discovered by Caldwell and Linder \cite{Caldwell}.  the $\omega_D-{\omega}^{\prime}_{D}$ plane presents the freezing region when both ${\omega}_{D}$  and ${\omega}^{\prime}_{D}$ are negative and the plane of $\omega_D<0$ and ${\omega}^{\prime}_{D}>0$ provides thawing region. In this case, the fixed point $\{\omega_D=-1,{\omega}^{\prime}_{D}=0\}$ shows the $\Lambda$CDM model. Here, prime denotes derivative
respect to $x=lna$.

At the classical level, the effects of perturbations on the stability of the DE model 
 is investigated by studying the sign of the sound
speed squared ($v_{s}^{2}$) \cite{Peebles}. the general form of $v_{s}^{2}$ is proposed as 
\begin{equation}\label{vs}
v_{s}^{2}=\frac{dP_D}{d\rho_D}=\frac{\dot{P}_D}{\dot{\rho}_D}=\dfrac{\rho_{D}}{\dot{\rho}_{D}}
\dot{\omega}_{D}+\omega_{D},
\end{equation}
where $v_{s}^{2}>0$  shows that the model is stable.
\section{NTADE in fractal cosmology: Non-interacting case}

In this section, we study the NTADE model in the framework of fractal spacetime by considering $b^2=0$. According with Ref.~\cite{TNADE}, by assuming  the conformal time $\eta$ as IR cutoff, the new Tsallis agegraphic energy density and its time derivative obtain as
\begin{eqnarray}\label{Tnage}
\rho_D=B{\eta}^{2\delta-4},
\end{eqnarray}
and
\begin{equation}\label{age1}
\dot{\rho}_D=\frac{2\delta-4}{a\eta}\rho_D,
\end{equation}
respectivily, where $d\eta=a dt$. Taking time derivative of the Friedmann equation 
(\ref{fractal}) and using Eqs.~(\ref{fractal1}) and
(\ref{age1}), we get
\begin{equation}\label{dotH}
\frac{\dot{H}}{H^2}=\frac{\frac{(\gamma-3)\Omega_{m_0} {H_0}^2
a^{\gamma-3}}{2H^2}+\frac{(\delta-2)B{\eta}^{2\delta-4}}{3a\eta
H^3}-\frac{\gamma^3 \omega
a^{-2\gamma}}{6}}{1-\gamma-\frac{\gamma^2 \omega
a^{-2\gamma}}{6}},
\end{equation}
where we take $\rho_{m_0}=3{H_0}^2 \Omega_{m_0}$.
Therefore, the deceleration parameter
$q(\equiv-1-\frac{\dot{H}}{H^2})$, what is interpreted as the rate of the acceleration or deceleration of the Universe expansion is found out as
\begin{equation}\label{qnage}
q=-1-\frac{\frac{(\gamma-3)\Omega_{m_0} {H_0}^2
a^{\gamma-3}}{2H^2}+\frac{(\delta-2)B{\eta}^{2\delta-4}}{3a\eta
H^3}-\frac{\gamma^3 \omega
a^{-2\gamma}}{6}}{1-\gamma-\frac{\gamma^2 \omega
a^{-2\gamma}}{6}}.
\end{equation}

Additionally, Eq.~(\ref{fractal}) helps us in finding the mathematical expression for the dimensionless NTADE density as
\begin{eqnarray}\label{Omega}
\Omega_D=\frac{B{\eta}^{2\delta-4}}{3H^2(1-\gamma-\frac{\gamma^2
\omega a^{-2\gamma}}{6})}.
\end{eqnarray}
We obtain the equation describing the evolution of DE density parameter by taking the time derivative of relation (\ref{Omega}),
\begin{equation}\label{dotOmega}
\dot{\Omega}_D=\Omega_D\left(\frac{\frac{2\delta-4}{a\eta}(1-\gamma-\frac{\gamma^2
\omega a^{-2\gamma}}{6})+H[2(1+q)(1-\gamma-\frac{\gamma^2 \omega
a^{-2\gamma}}{6})-\frac{1}{3}\gamma^3 \omega
a^{-2\gamma}]}{1-\gamma-\frac{\gamma^2 \omega
a^{-2\gamma}}{6}}\right).
\end{equation}
Also, if we insert Eq.~(\ref{age1}) in Eq.~(\ref{fractal2}), we obtain the equation of state (EOS) parameter as
\begin{eqnarray}\label{EoSna}
\omega_D=-1-\frac{2\delta-4}{(3-\gamma)a\eta H}.
\end{eqnarray}
In order to study the stability of the NTADE model in the present scenario, we take the time derivative of Eq.~(\ref{EoSna}) and insert the
result in Eq.~(\ref{vs}), which finally leads to
\begin{equation}\label{vs1}
v_{s}^{2}=\frac{2-\gamma-\frac{\gamma^3
\omega}{6a^{2\gamma}(\gamma-1)+\gamma^2
\omega}-\frac{5-2\delta}{a\eta
H}}{\gamma-3}+\frac{2a^{-1+2\gamma}B(\delta-2)\eta^{-5+2\delta}}{(\gamma-3)H^3(6a^{2\gamma}(\gamma-1)+\gamma^2
\omega)}\\+\frac{3a^{3(\gamma-1)}H_0^2
\Omega_{m_0}}{H^2(6a^{2\gamma}(\gamma-1)+\gamma^2 \omega)},
\end{equation}
 for the non-interacting case.

\begin{figure}[htp]
\begin{center}
\includegraphics[width=8cm]{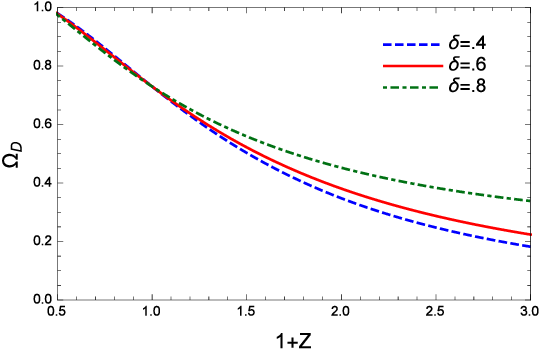}
\includegraphics[width=8cm]{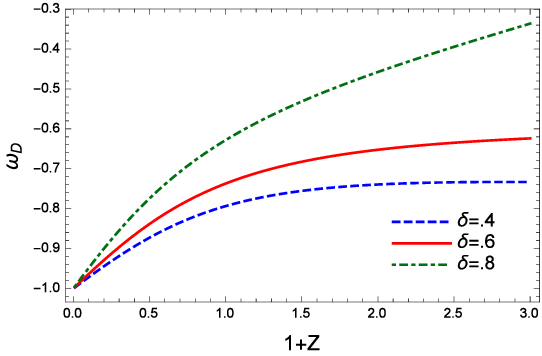}
\caption{Left gragh: The evolution of $\Omega_D$ versus redshift parameter $z$ for
 non-interacting NTADE in Fractal cosmology. Right gragh: The evolution of $\omega_D$ versus redshift parameter $z$ for
 non-interacting NTADE in Fractal cosmology. Here, we have taken
$\Omega_D(z=0)=0.73$, $H(z=0)=74$, $\omega=.3$, $\gamma=.1$ and $B=2.4$
}\label{Omega-z1}
\end{center}
\end{figure}


\begin{figure}[htp]
\begin{center}
\includegraphics[width=8cm]{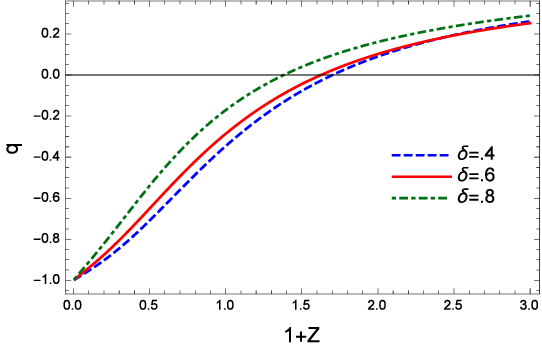}
\includegraphics[width=8cm]{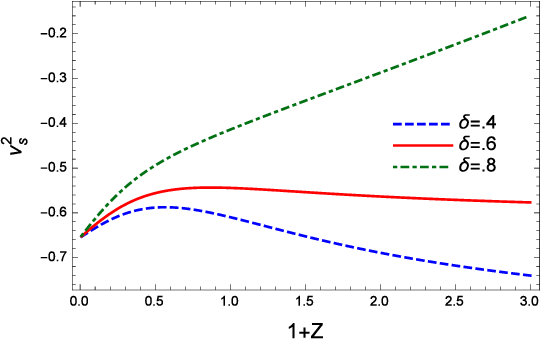}
\caption{Left gragh: The evolution of the deceleration parameter $q$ versus redshift parameter $z$ for
 non-interacting NTADE in Fractal cosmology. Right gragh: The evolution of  the squared of sound speed $v_s^2 $ versus redshift parameter $z$ for
 non-interacting NTADE in Fractal cosmology. Here, we have taken
$\Omega_D(z=0)=0.73$, $H(z=0)=74$, $\omega=.3$, $\gamma=.1$ and $B=2.4$}\label{q-z1}
\end{center}
\end{figure}

\begin{figure}[htp]
\begin{center}
\includegraphics[width=6cm]{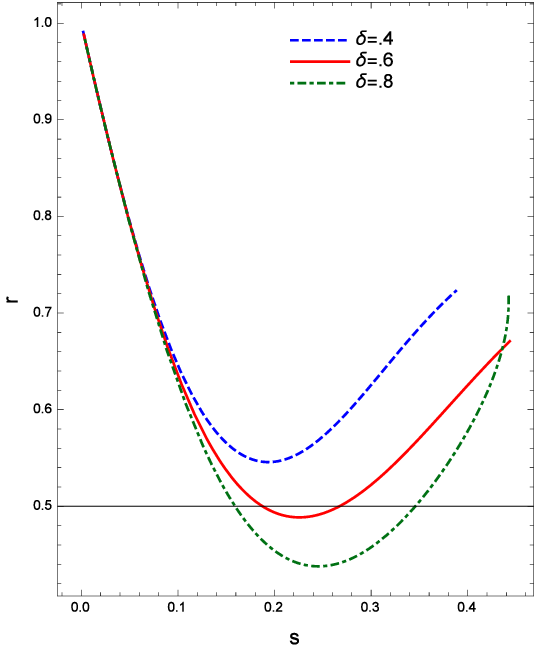}
\includegraphics[width=8cm]{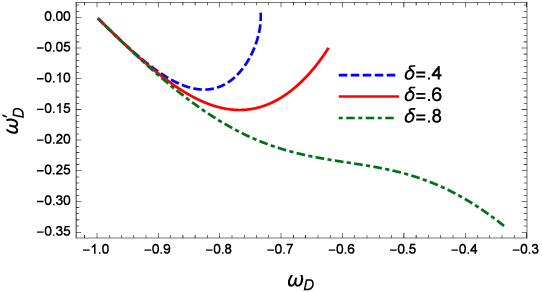}
\caption{Left gragh: The evolution of the statefinder parameter $r$ versus $s$ for
 non-interacting NTADE in Fractal cosmology. Right gragh: The $\omega_D-{\omega}^{\prime}_{D}$ diagram for
non-interacting NTADE in Fractal cosmology. Here, we have taken
$\Omega_D(z=0)=0.73$, $H(z=0)=74$, $\omega=.3$, $\gamma=.1$ and $B=2.4$}\label{rs-z1}
\end{center}
\end{figure}



To obtain the values of $\{r,s\}$ plane, we insert Eq.~(\ref{qnage}) and its time
derivative in relations (\ref{rr2}) and (\ref{statefinder}), after some calculations, we have 
\begin{align}\label{r1}
r=-\frac{1}{a^4 \eta ^{10} (6 a^{2 \gamma } (-1+\gamma )+\gamma ^2 \omega )^2 H^6}\Bigg(4 a^{2+4 \gamma
} B^2 (-2+\delta )^2 \eta ^{4 \delta }+\eta ^4 H \Bigg[15 a^{1+3 \gamma } {H_0}^2 \Omega_{m_0} (-3+\gamma ) \gamma ^3 \eta ^6 \omega
 H^3\nonumber\\
-a^4 \gamma ^4 (1+\gamma ) (1+2 \gamma ) \eta ^6 \omega ^2 H^5
+6 a^{5 \gamma } {H_0}^2 \Omega_{m_0} (-3+\gamma ) \eta  (B (-2+\delta ) \eta ^{2 \delta }+3 a (-1+\gamma ) \gamma  \eta ^5 H^3)~~~~~~~~~~~~\nonumber\\
+12 a^{2+4 \gamma } (-1+\gamma ) H (-3 a^2 (-1+\gamma ) \eta ^6 H^4+B (-2+\delta ) \eta ^{2 \delta } (-5+2 \delta +2 a \eta
 H))~~~~~~~~~~~~~~~~~~~~~~~~~~~~~~~\nonumber\\
+2 a^{2+2 \gamma } \gamma ^2 \omega  H \Big(3 a^2 (-2+\gamma ) (-1+\gamma ) (1+2 \gamma ) \eta ^6 H^4+B (-2+\delta
) \eta ^{2 \delta } (-5+2 \delta +a (2+3 \gamma ) \eta  H)\Big)\Bigg]\Bigg).~~~~
\end{align}

\begin{align}\label{s1}
s=2\Bigg(4 a^{2+4 \gamma } B^2 (-2+\delta )^2 \eta ^{4 \delta}+
\eta ^4 H
\Bigg[6 a^{5 \gamma } B {H_0}^2 \Omega_{m_0} (-3+\gamma ) (-2+\delta ) \eta ^{1+2 \delta }~~~~~~~~~~~~~~~~~~~~~~~~~~~~~~~~~~~~~~~~~~~~~~~~~~~~~~~~~~~~~~~~\nonumber\\
+aH\Big(\gamma  \eta ^6 H^2 (3 a^{3 \gamma } {H_0}^2 \Omega_{m_0} (-3+\gamma ) (6 a^{2 \gamma } (-1+\gamma )+5 \gamma ^2
\omega)+a^3 \gamma ^2 \omega  (6 a^{2 \gamma } (-1+\gamma ) (-3+2 \gamma )-\gamma ^2 (3+2 \gamma ) \omega) H^2)~~~~~~~~~~~~~~~~\nonumber\\
+2 a^{1+2 \gamma } B (-2+\delta ) \eta ^{2 \delta } \Big[6 a^{2 \gamma } (-1+\gamma ) (-5+2 \delta +2 a \eta
 H)+\gamma ^2 \omega  (-5+2 \delta +a (2+3 \gamma ) \eta  H)\Big]\Big)\Bigg]\Bigg)~~~~~~~~~~~~~~~~~~~~~~~~~~~~~~~~~~~~~~~~\nonumber\\  \Bigg/\Bigg(3 a \eta ^5 (6 a^{2 \gamma } (-1+\gamma
)+\gamma ^2 \omega)
H^3 ~~~~~~~~~~~~~~~~~~~~~~~~~~~~~~~~~~~~~~~~~~~~~~~~~~~~~~~~~~~~~~~~~~~~~~~~~~~~~~~~~~~~~~~~~~~~~~~~~~~~~~~~~~~~~~~~~~~~\nonumber\\\times\Big(-4 a^{2+2 \gamma}B (-2+\delta ) \eta ^{2 \delta }+\eta ^5 H(-6 a^{3 \gamma } {H_0}^2 \Omega_{m_0} (-3+\gamma
)+a^3 (18 a^{2 \gamma } (-1+\gamma )+\gamma ^2 (3+2 \gamma ) \omega ) H^2)\Big)\Bigg).~~~~~~~~~~~~~~~~~~~~~~~~~~
\end{align}
Also, we obtain the evolution of EoS parameter as
\begin{align}\label{wp1}
{\omega}^{\prime}_{D}= \frac{1}{a^4 (-3+\gamma ) \eta ^6(6 a^{2 \gamma } (-1+\gamma )+\gamma ^2 \omega ) H^4}\Bigg(2 (-2+\delta ) \Bigg[2 a^{2+2 \gamma } B (-2+\delta ) \eta ^{2 \delta }+\eta ^4H~~~~~~~~~~~~ \nonumber\\ ~~~~~~~\times\Big(3 a^{3 \gamma
} {H_0}^2 \Omega_{m_0} (-3+\gamma ) \eta +a^2 H\Big[-6 a^{2 \gamma } (-1+\gamma ) (1+a \eta  H)-\gamma ^2 \omega  (1+a (1+\gamma ) \eta
 H)\Big]\Big)\Bigg]\Bigg).
\end{align}

In Fig.~\ref{Omega-z1}, left gragh, we plot the evolution of $\Omega_D$ versus redshift parameter $z$ for different values of $\delta$ parameter. We observe the dominance of dark energy increases as we increase the value of $\delta$ in the past, while the Universe is completely dominated by dark energy for various values of $\delta$ at the future. The right gragh of Fig.~\ref{Omega-z1} shows the evolution of the EoS parameter $\omega_D$ versus redshift $z$ for the different values of $\delta$. This depicts that the EoS parameter lies in quintessence region and converges to the cosmological constant $\omega_D=-1$ at future. The behavior of the deceleration parameter $q$ and square of the sound speed parameter ${v}^{2}_{s}$ is shown in the left and right gragh of Fig.~\ref{q-z1} respectively. The left gragh of Fig.~\ref{q-z1} indicates that as that we decerase value of $\delta$, the transition from early decelerated phase to present accelerating phase occures faster. From right gragh, we observe that model is not stable for all values of $\delta$. The evolutionary trajectories of the statefinder pair $\{r,s\}$ are shown in the left gragh of Fig.~\ref{rs-z1} for various values of $\delta$ parameter. We observe the trajectories begin in the quintessence region during an early time and as the Universe is expanding end at $\Lambda$CDM 
 $\{r,s\}=\{1,0\}$ in late time. The $\omega_D-{\omega_D}^{\prime}$ trajectories have been ploted in the right gragh of Fig.~\ref{rs-z1}. As shown in this figure, the EoS parameter $\omega_D$ as well as its evolution measured by ${\omega_D}^{\prime}$ lie in the negative region or in the other words, we encounter the freezing region which shows the cosmos expands with a more accelerated rate than the thawing region.


\section{NTADE in fractal cosmology: Interacting case}

In addition to the observational evidence that supports the interaction between DM and DE \cite{Bertolami}, response to this question that why are the matter and dark energy densities the same order?, called the coincidence problem, is a main reason to assume the interaction between the dark sectors of the Universe. At first glance, the unknown nature of DE as well as DM may make the basic problem for the choice of interaction term but, in view of the continuty equations it must be a function of the product of energy density and a term with unit of the inverse of time namely, Hubble parameter. Hence, different forms for $Q$ can be proposed \cite{Bolotin}. Here, we choose $Q=3b^2 H \rho_m$ term.
Similar to the previous section, taking the time derivative of Eq. (\ref{fractal}) along with combining the result with the continuity equations, we arrive at
\begin{equation}\label{dotH1}
\frac{\dot{H}}{H^2}=\frac{\frac{(\gamma-3+3b^2)\Omega_{m_0} {H_0}^2 a^{\gamma-3+3b^2}}{2H^2}+\frac{(\delta-2)B{\eta}^{2\delta-4}}{3a\eta H^3}-\frac{\gamma^3 \omega a^{-2\gamma}}{6}}{1-\gamma-\frac{\gamma^2 \omega a^{-2\gamma}}{6}},
\end{equation}
which also leads to the expression for the deceleration parameter as

\begin{equation}\label{qnage1}
q=-1-\frac{\frac{(\gamma-3+3b^2)\Omega_{m_0} {H_0}^2 a^{\gamma-3+3b^2}}{2H^2}+\frac{(\delta-2)B{\eta}^{2\delta-4}}{3a\eta H^3}-\frac{\gamma^3 \omega a^{-2\gamma}}{6}}{1-\gamma-\frac{\gamma^2 \omega a^{-2\gamma}}{6}}.
\end{equation}
By combining Eqs.~(\ref{fractal2}) and ~(\ref{age1}), we extract the expression of EoS parameter as
\begin{eqnarray}\label{EoSna1}
\omega_D=-1-\frac{2\delta-4}{(3-\gamma)a\eta H}-\frac{9b^2 H_0^2 \Omega_{m_0} a^{3b^2-3+\gamma}}{B(3-\gamma){\eta}^{2\delta-4}}.
\end{eqnarray}
The time derivative of Eq.~(\ref{EoSna1}) along with relation (\ref{vs}) gives the mathematical expression for the squared speed of sound as
\begin{align}
v_{s}^{2}=\frac{2-\gamma-\frac{\gamma^3 \omega}{6 a^{2\gamma}(\gamma-1)+\gamma^2 \omega}-\frac{5-2\delta}{a\eta H}}{\gamma-3}
+\frac{2a^{-1+2\gamma}B(\delta-2)\eta^{-5+2\delta}}{(\gamma-3)H^3(6a^{2\gamma}(\gamma-1)+\gamma^2 \omega)}+\frac{3a^{3(\gamma-1+b^2)}H_0^2 \Omega_{m_0}(-3+3b^2+\gamma)}{H^2(6a^{2\gamma}(\gamma-1)+\gamma^2 \omega)(\gamma-3)}\nonumber\\
+\frac{9b^2a^{-2+\gamma+3b^2}H_0^2 \Omega_{m_0}(-3+3b^2+\gamma)\eta^{5-2\delta} H}{2B(\gamma-3)(\delta-2)}.\label{vs2}~~~~~~~~~~~~~~~~~~~~~~~~~~~~~~~~~~~~~~~~~~~~~~~~~~~~~~~~~~~~~~~~~~~~~~~~
\end{align}


\begin{figure}[htp]
\begin{center}
\includegraphics[width=8cm]{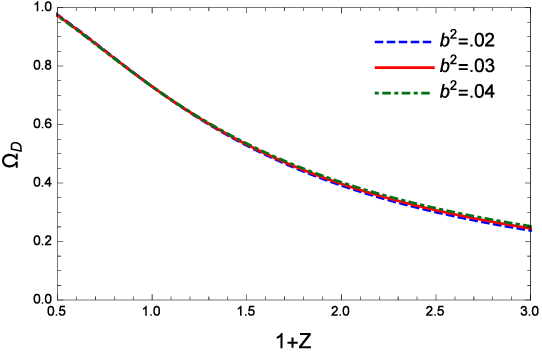}
\includegraphics[width=8cm]{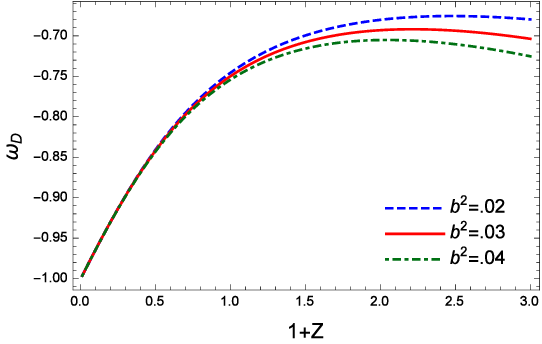}
\caption{Left gragh: The evolution of $\Omega_D$ versus redshift parameter $z$ for
 interacting NTADE in Fractal cosmology. Right gragh: The evolution of $\omega_D$ versus redshift parameter $z$ for
 interacting NTADE in Fractal cosmology. Here, we have taken
$\Omega_D(z=0)=0.73$, $H(z=0)=74$, $\omega=.3$, $\gamma=.1$, $B=2.4$ and $\delta=.6$
}\label{Omega-z2}
\end{center}
\end{figure}


\begin{figure}[htp]
\begin{center}
\includegraphics[width=8cm]{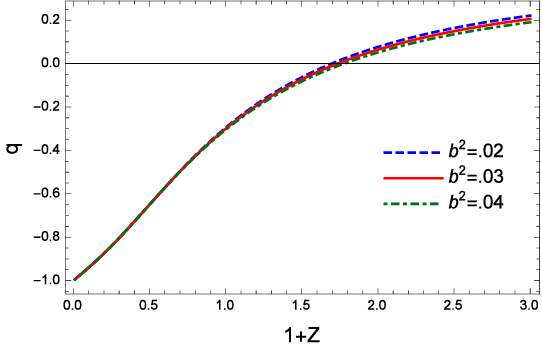}
\includegraphics[width=8cm]{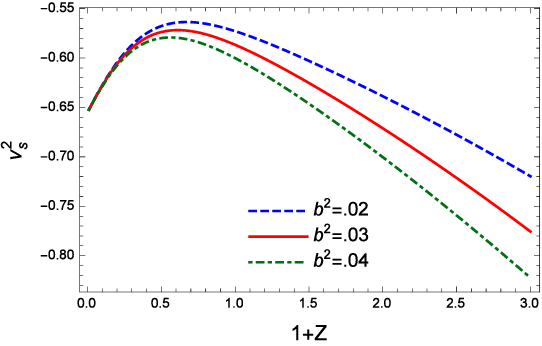}
\caption{Left gragh: The evolution of the deceleration parameter $q$ versus redshift parameter $z$ for
 interacting NTADE in Fractal cosmology. Right gragh: The evolution of  the squared of sound speed $v_s^2 $ versus redshift parameter $z$ for
 interacting NTADE in Fractal cosmology. Here, we have taken
$\Omega_D(z=0)=0.73$, $H(z=0)=74$, $\omega=.3$, $\gamma=.1$, $B=2.4$ and $\delta=.6$}\label{q-z2}
\end{center}
\end{figure}


\begin{figure}[htp]
\begin{center}
\includegraphics[width=8cm]{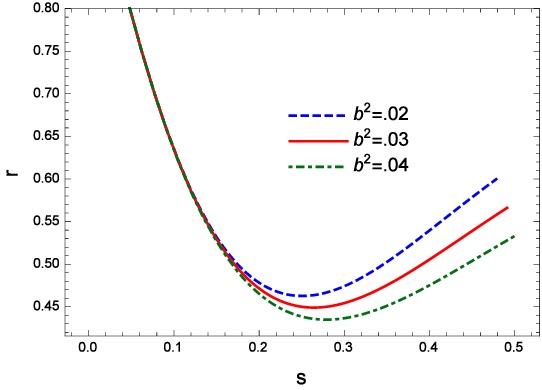}
\includegraphics[width=8cm]{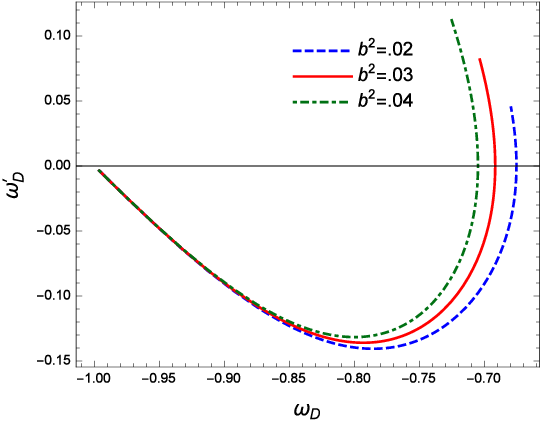}
\caption{Left gragh: The evolution of the statefinder parameter $r$ versus $s$ for
 interacting NTADE in Fractal cosmology. Right gragh: The $\omega_D-{\omega}^{\prime}_{D}$ diagram for
 interacting NTADE in Fractal cosmology. Here, we have taken
$\Omega_D(z=0)=0.73$, $H(z=0)=74$, $\omega=.3$, $\gamma=.1$, $B=2.4$ and $\delta=.6$}\label{rs-z2}
\end{center}
\end{figure}



\begin{figure}[htp]
\begin{center}
\includegraphics[width=8cm]{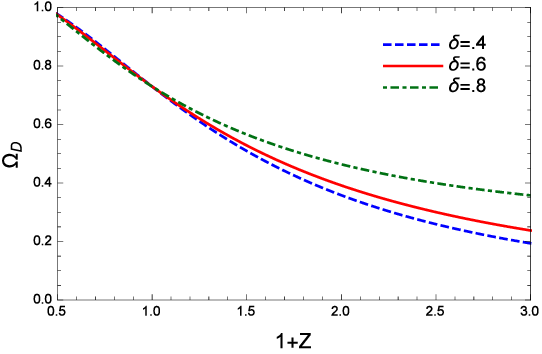}
\includegraphics[width=8cm]{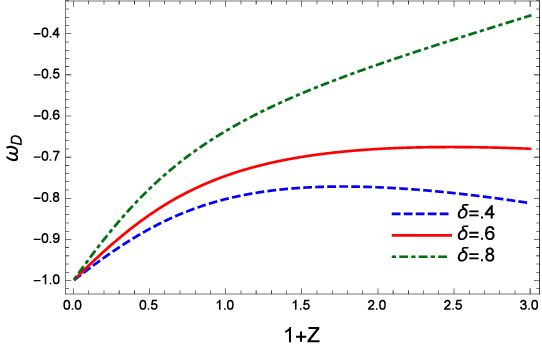}
\caption{Left gragh: The evolution of $\Omega_D$ versus redshift parameter $z$ for
 interacting NTADE in Fractal cosmology. Right gragh: The evolution of $\omega_D$ versus redshift parameter $z$ for
 interacting NTADE in Fractal cosmology. Here, we have taken
$\Omega_D(z=0)=0.73$, $H(z=0)=74$, $\omega=.3$, $\gamma=.1$, $B=2.4$ and $b^2=.02$
}\label{Omega-z3}
\end{center}
\end{figure}


\begin{figure}[htp]
\begin{center}
\includegraphics[width=8cm]{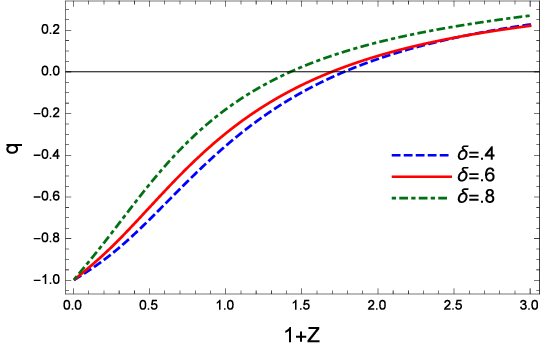}
\includegraphics[width=8cm]{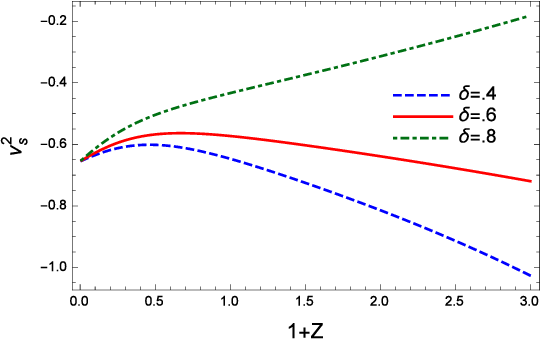}
\caption{Left gragh: The evolution of the deceleration parameter $q$ versus redshift parameter $z$ for
 interacting NTADE in Fractal cosmology. Right gragh: The evolution of  the squared of sound speed $v_s^2 $ versus redshift parameter $z$ for
 interacting NTADE in Fractal cosmology. Here, we have taken
$\Omega_D(z=0)=0.73$, $H(z=0)=74$, $\omega=.3$, $\gamma=.1$, $B=2.4$ and $b^2=.02$}\label{q-z3}
\end{center}
\end{figure}


\begin{figure}[htp]
\begin{center}
\includegraphics[width=6cm]{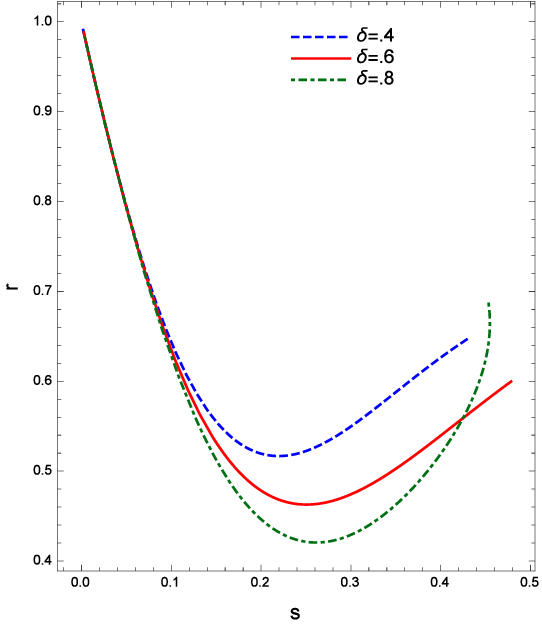}
\includegraphics[width=8cm]{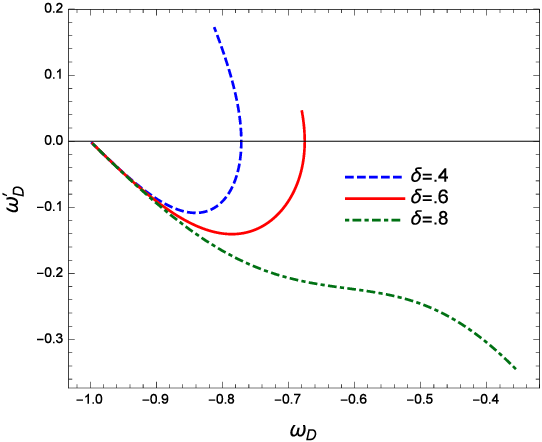}
\caption{Left gragh: The evolution of the statefinder parameter $r$ versus $s$ for
 interacting NTADE in Fractal cosmology. Right gragh: The $\omega_D-{\omega}^{\prime}_{D}$ diagram for
 interacting NTADE in Fractal cosmology. Here, we have taken
$\Omega_D(z=0)=0.73$, $H(z=0)=74$, $\omega=.3$, $\gamma=.1$, $B=2.4$ and $b^2=.02$}\label{rs-z3}
\end{center}
\end{figure}



To determine the $\omega_D-{\omega}^{\prime}_{D}$ plane, we take derivative of $\omega_D$ respect to $x=lna$
\begin{align}\label{wp2}
{\omega}^{\prime}_{D}=\frac{1}{a^4 (-3+\gamma ) \eta ^2}\Bigg(-\frac{9 a^{1+3 b^2+\gamma} b^2 {H_0}^2 \Omega_{m_0} (-3+3 b^2+\gamma ) \eta ^{6-2 \delta}}{B}+\frac{2 a^2 (-2+\delta )}{H^2}+\frac{2
a^3 (-2+\delta ) \eta }{H}~~\nonumber\\+\frac{18 a^{3 b^2+\gamma } b^2 {H_0}^2 \Omega_{m_0} (-2+\delta ) \eta ^{5-2 \delta }}{BH}\Bigg)~~~~~~~~~~~~~~~~~~~~~~~~~~~~~~~~~~~~~~~~~~~~~~~~~~~~~~~~~~~~~~~~~~~~~~~~~~~~\nonumber\\
+\frac{1}{a^4 (-3+\gamma ) \eta ^2}\Bigg(\Bigg[2 (-2+\delta
) \Big(-2 a^{2+2 \gamma } B (-2+\delta ) \eta ^{2 \delta}+\eta ^5 H(-3 a^{3 (b^2+\gamma )}{H_0}^2 \Omega_{m_0} (-3+3 b^2+\gamma
)\nonumber\\~~~~~~~~~~~+a^3 \gamma ^3 \omega  H^2)\Big)\Bigg] \Bigg/\Big(\eta ^4 (6 a^{2 \gamma } (-1+\gamma )+\gamma ^2 \omega) H^4\Big)\Bigg).~~~~~~~~~~~~~~~~~~~~~~~~~~~~~~~~~~~~~~~~~~~~~~~~~~~~~~~~~~~
\end{align}
Finally, we can calculate the statefinder parameters $r$ and $s$ by using the relations (\ref{rr2}) and (\ref{statefinder})
\begin{align}\label{r2}
r=-\frac{1}{a^4 \eta ^{10} (6 a^{2 \gamma } (-1+\gamma )+\gamma ^2 \omega )^2 H^6}\Bigg(4 a^{2+4 \gamma
} B^2 (-2+\delta )^2 \eta ^{4 \delta }+\eta ^4 H \Bigg[3 a^{1+3b^2+3 \gamma } {H_0}^2 \Omega_{m_0} \gamma ^2(-3+3b^2+\gamma )~~~~~~~~~~~~~~~~~~~~~~~~~~~\nonumber\\\times (3b^2+5\gamma) \eta ^6 \omega
 H^3
-a^4 \gamma ^4 (1+\gamma ) (1+2 \gamma ) \eta ^6 \omega ^2 H^5
+6 a^{3b^2+5 \gamma } {H_0}^2 \Omega_{m_0} (-3+3b^2+\gamma ) \eta  (B (-2+\delta ) \eta ^{2 \delta }+3 a (-1+\gamma )~~~~~~~~~~~~~\nonumber\\\times (3b^2+\gamma)  \eta ^5 H^3)
+12 a^{2+4 \gamma } (-1+\gamma ) H (-3 a^2 (-1+\gamma ) \eta ^6 H^4+B (-2+\delta ) \eta ^{2 \delta } (-5+2 \delta +2 a \eta
 H))~~~~~~~~~~~~~~~~~~~~~~~~~~~~~~~~~~~~~~\nonumber\\
+2 a^{2+2 \gamma } \gamma ^2 \omega  H \Big(3 a^2 (-2+\gamma ) (-1+\gamma ) (1+2 \gamma ) \eta ^6 H^4+B (-2+\delta
) \eta ^{2 \delta } (-5+2 \delta +a (2+3 \gamma ) \eta  H)\Big)\Bigg]\Bigg).~~~~~~~~~~~~~~~~~~~~~~~~~~~~~~~~~~~
\end{align}

\begin{align}\label{s2}
s=2\Bigg(4 a^{2+4 \gamma }B^2 (-2+\delta )^2 \eta ^{4 \delta }+
\eta ^4 H \Bigg[3 a^{1+3 b^2+3 \gamma } {H_0}^2 \Omega_{m_0} \gamma ^2 (-3+3 b^2+\gamma ) (3 b^2+5 \gamma ) \eta ^6 \omega  H^3-a^4 \gamma
^5 (3+2 \gamma ) \eta ^6 \omega ^2H^5~~~~~~~~~~~~~~~~~~~~~~~~~~~~~~~~~~~~~~~~~~~~~~~~~~~~~~~~~~~~~~~~~~~~~~~~~~~~~~~~~~~~~~~~~~~~~~~~~~~~~~~~~~~~~~~~~~~~~~~~~~~~~~~~~~~~~~~~~~~~~~~~~~~~~~~~~~~~~~~~~~~~~~~~~~~~~~~~~~~~~~~~~~~~~~~\nonumber\\
+12 a^{2+4 \gamma } B (-1+\gamma ) (-2+\delta ) \eta ^{2 \delta } H (-5+2 \delta +2 a \eta  H)+6 a^{3 b^2+5 \gamma } {H_0}^2 \Omega_{m_0} (-3+3 b^2+\gamma ) \eta ~~~~~~~~~~~~~~~~~~~~~~~~~~~~~~~~~~~~~~~~~~~~~~~~~~~~~~~~~~~~~~~~~~~~~~~~~~~~~~~~~~~~~~~~~~~~~~~~~~~~~~~~~~~~~~~~~~~~~~~~~~~~~~~~~~~~~~~~~~~~~~~~~~~~~~~~~~~~~~~~~~~~~~~~~~~~~~~~~~~~~~~~~~~~~~~~~~~~~~~~~~~~~~~~~~~~~~~~~~~~~ \nonumber\\\times(B (-2+\delta ) \eta ^{2 \delta } +3a (-1+\gamma ) (3 b^2+\gamma ) \eta ^5 H^3)
+2 a^{2+2 \gamma } \gamma ^2 \omega  H \Big(3 a^2 (-1+\gamma ) \gamma  (-3+2 \gamma ) \eta ^6 H^4~~~~~~~~~~~~~~~~~~~~~~~~~~~~~~~~~~~~~~~~~~~~~~~~~~~~~~~~~~~~~~~~~~~~~~~~~~~~~~~~~~~~~~~~~~~~~~~~~~~~~~~~~~~~~~~~~~~~~~~~~~~~~~~~~~~~~~~~~~~~~~~~~~~~~~~~~~~~~~~~~~~~~~~~~~~~~~~~~~~~~~~~~~~~~~~~~~~~~~~~~~~~~~~~~~~~~~~~\nonumber\\+B (-2+\delta
) \eta ^{2 \delta } (-5+2 \delta+a (2+3 \gamma ) \eta  H)\Big)\Bigg]\Bigg)~~~~~~~~~~~~~~~~~~~~~~~~~~~~~~~~~~~~~~~~~~~~~~~~~~~~~~~~~~~~~~~~~~~~~~~~~~~~~~~~~~~~~~~~~~~~~~~~~~~~~~~~~~~~~~~~~~~~~~~~~~~~~~~~~~~~~~~~~~~~~~~~~~~~~~~~~~~~~~~~~~~~~~~~~~~~~~~~~~~~~~~~~~~~~~~~~~~~~~~~~~~~~~~~~~~~~~~~~~~~~~~~~~~~~~~~~~~~~~~~~~~~~~~~~~~~~~~~~~~~~~~~~~~~~~~~~~~~~~\nonumber\\ \Bigg/
\Bigg(3 a \eta ^5 (6 a^{2 \gamma } (-1+\gamma )+\gamma ^2 \omega) H^3 \Bigg[-4 a^{2+2 \gamma } B(-2+\delta ) \eta ^{2
\delta }+\eta ^5 H \Big(-6 a^{3 (b^2+\gamma )} {H_0}^2 \Omega_{m_0} (-3+3 b^2+\gamma )~~~~~~~~~~~~~~~~~~~~~~~~~~~~~~~~~~~~~~~~~~~~~~~~~~~~~~~~~~~~~~~~~~~~~~~~~~~~~~~~~~~~~~~~~~~~~~~~~~~~~~~~~~~~~~~~~~~~~~~~~~~~~~~~~~~~~~~~~~~~~~~~~~~~~~~~~~~~~~~~~~~~~~~~~~~~~~~~~~~~~~~~~~~~~~~~~~~~~~~~~~~~\nonumber\\+a^3 (18 a^{2 \gamma } (-1+\gamma )+\gamma ^2 (3+2 \gamma
) \omega )H^2\Big)\Bigg]\Bigg).~~~~~~~~~~~~~~~~~~~~~~~~~~~~~~~~~~~~~~~~~~~~~~~~~~~~~~~~~~~~~~~~~~~~~~~~~~~~~~~~~~~~~~~~~~~~~~~~~~~~~~~~~~~~~~~~~~~~~~~~~~~~~~~~~~~~~~~~~~~~~~~~~~~~~~~~~~~~~~~~~~~~~~~~~~~~~~~~~~~~~~~~~~~~~~~~~~~~~~~~~~~~~~~~~~~~~~~~~~~~~~~~~~~~~~~~~~~~~~~~~~~~~~~~~~~~~~~~~~~~~~~~~~~~~~~~~~~~~~~~~
\end{align}

The model behavior has been depicted in Figs.~\ref{Omega-z2}-\ref{rs-z2} for interacting case by assuming fixed value of $\delta$ and in Figs.~\ref{Omega-z3}-\ref{rs-z3} for fixed value of $b^2$ respectively. We observe from Fig.~\ref{Omega-z2} that, response of the evolution of $\Omega_D$ is same for all values of interaction term $b^2$, as well as, the EoS parameter wouldn\textquoteright t like to enter phantom region. The gragh of Fig.~\ref{q-z2} show that $q$ changes its sign from positive to negative without sensitivity to different values of $b^2$ as well as, the sign of $v_s^2 $ is negative which supports instability of model. The left gragh of Fig.~\ref{rs-z2}, indicates the evolutionary trajectory in $\{r-s\}$ plane for various values of interaction term $b^2$. This gragh has a behavior similar to quintessence model, like non-interacting case in Fig.~\ref{rs-z1}, although the $\omega_D-{\omega}^{\prime}_{D}$ trajectories depict both evolutionary zones of thawing and freezing unlike to Fig.~\ref{rs-z1}. As mentioned earlier, we plot the behavior of cosmological parameters in Figs.~\ref{Omega-z3}-\ref{rs-z3} for different values of $\delta$ and fixed value of $b^2$. We observe that the graghs are similar to non-interacting case. It shows $\delta$ parameter affects this dark energy model  more than the interaction term $b^2$.

\section{Closing remarks}
Reconciling general relativity with quantum mechanics, called quantum gravity, is a priority  for many theoretical physicists. the theories related to this field give a hint that dimension of spacetime changes with tha scale of the Universe in other words, ordinary geometry is not adequate to describe the microscopic texture of spacetime. This is a characteristic of fractals and refers to the fact that, the  application of quantum mechanics to spacetime leads to a fractal geometry.

 Some of dark energy models are based on the entropy relation in such a way that, each modification to the system entropy may change them. Based on this, when Tsallis introduced a non-extensive entropy for systems with a long range interaction, its applying made new dark energy models. 
In the present paper, we studied the NTADE model in the
framework of the fractal cosmology. It is observed that the EoS parameter depicts quintessence-like behavior for both interacting and non-interacting cases, similar to the $r-s$ analysis. Also, this scenario cannot obtain a sufficient condition for the classical stability. Here, we encounter the freezing region for $\omega_D-{\omega}^{\prime}_{D}$ plane which is the best match for scenario of current acceleration.


\acknowledgments{The author thanks Salman Farsi University of Kazerun Research Council .
}

\end{document}